\documentclass[journal = nalefd, manuscript = letter, layout = traditional]{achemso}

\usepackage{graphicx}
\usepackage{multirow}
\usepackage{color}
\usepackage{dcolumn}
\usepackage{bm}
\usepackage{subfigure}
\usepackage{rotating}
\usepackage{amsmath}

\author{Taras Plakhotnik}
\email{taras@physics.uq.edu.au}
\affiliation[University of Queensland]{School of Mathematics and Physics, The University of Queensland, Queensland 4072, Australia}
\author{Marcus W. Doherty}
\affiliation[Australian National University]{Laser Physics Centre, Research School of Physics and Engineering, Australian National University, Australian Capital Territory 0200, Australia}
\author{Jared H. Cole}
\affiliation[RMIT University]{Chemical and Quantum Physics, School of Applied Sciences, RMIT University, Victoria 3001, Australia}
\author{Robert Chapman}
\affiliation[University of Queensland]{School of Mathematics and Physics, The University of Queensland, Queensland 4072, Australia}
\author{Neil B. Manson}
\affiliation[Australian National University]{Laser Physics Centre, Research School of Physics and Engineering, Australian National University, Australian Capital Territory 0200, Australia}

\title{All-optical thermometry and the thermal properties of the optically detected spin resonances of the NV$^-$ center in nano-diamond}
\keywords{Nitrogen-vacancy center, diamond, electron spin resonance, Debye-Waller factor, thermometry}

\begin{document}

\begin{abstract}
The negatively-charged nitrogen-vacancy (NV$^-$) center in diamond is at the frontier of quantum nano-metrology and bio-sensing. Recent attention has focused on the application of high-sensitivity thermometry using the spin resonances of NV$^-$ centers in nano-diamond to sub-cellular biological and biomedical research. Here, we report a comprehensive investigation of the thermal properties of the center's spin resonances and demonstrate an alternate all-optical NV$^-$ thermometry technique that exploits the temperature dependence of the center's optical Debye-Waller factor.
\end{abstract}

\begin{tocentry}
\includegraphics[width=0.6\columnwidth] {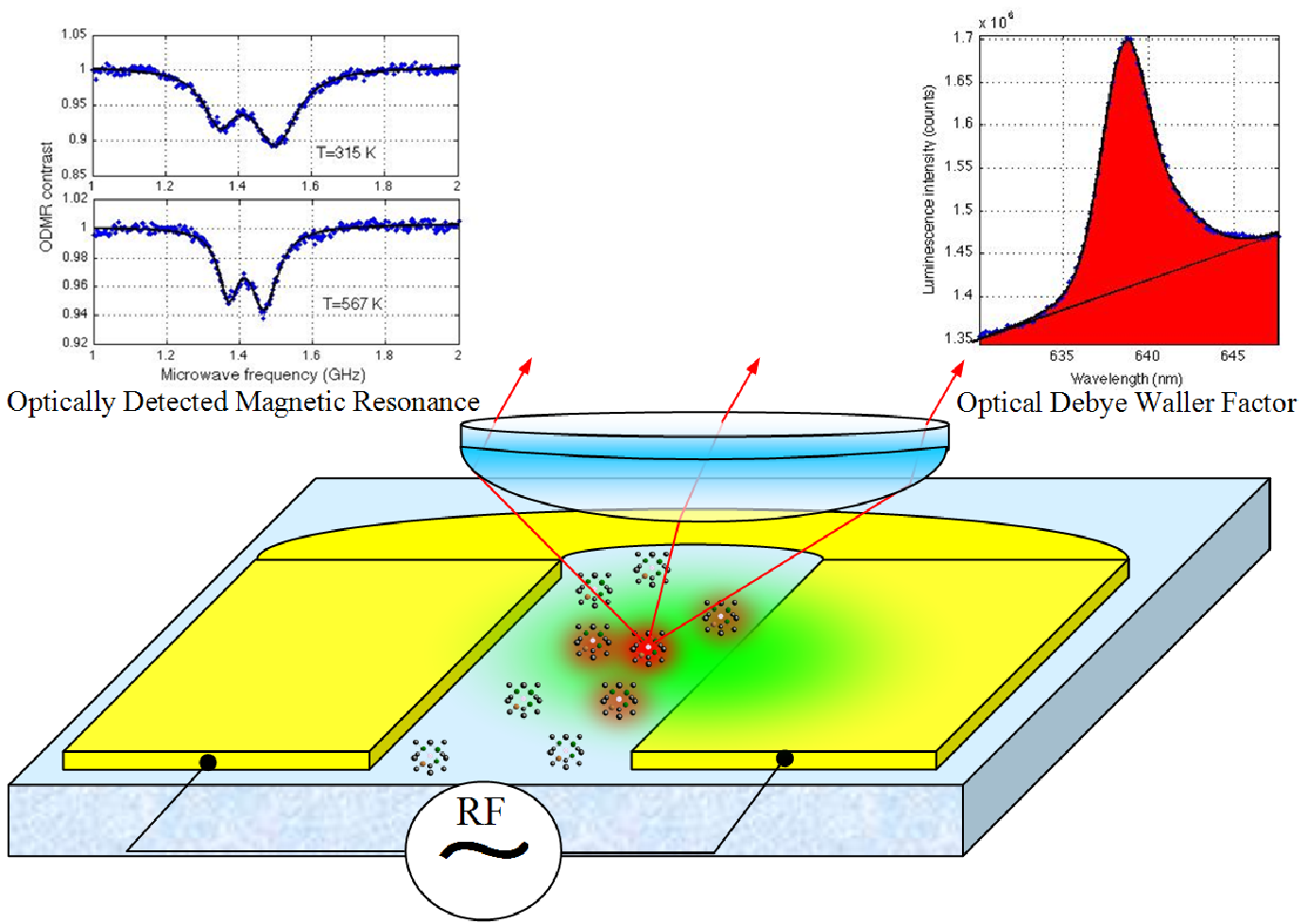}
\end{tocentry}

The negatively-charged nitrogen-vacancy (NV$^-$) center in diamond \cite{review} is an important defect system for a range of quantum technologies. The center has been key to several advances in quantum information processing, including the realization of solid-state spin quantum registers and elements of quantum networks \cite{qip1,qip2,qip3,qip4,qip5}.  The center has also attracted significant attention due its potential as a high-sensitivity nanoscale quantum sensor that can operate in ambient and extreme conditions. Indeed, the NV$^-$ center has been employed in many impressive demonstrations of nano- magnetometry \cite{mag1,mag2,mag3,mag4}, electrometry \cite{efield,dolde14}, piezometry \cite{doherty14} and thermometry \cite{toyli12,toyli13,neumann13,kucsko13,plak10}. Furthermore, the center may be incorporated into nano-diamonds, whose chemical inertness and biocompatibility, allow sensing techniques to be performed within living cells \cite{bioimag}. Recently, the center's bio-sensing applications were expanded with the demonstration of sub-cellular temperature gradient mapping and control using NV$^-$ nano-diamond and gold nano-particles within a human embryonic fibroblast \cite{kucsko13}. This exciting demonstration can be potentially extended to \textit{in vivo} thermometry and thermoablative therapy \cite{kucsko13}.

Each of the center's metrology applications either directly employ its electron spin or a nuclear spin that is coupled to its electron spin as a quantum sensor. High sensitivity is principally achieved by the unique combination of the exceptional electromagnetic, mechanical and thermal properties of diamond and the long-lived coherence of the center's electron spin, which persists in ambient and extreme conditions \cite{toyli12,balasubramanian09,pham12}. Nanoscale sensing is enabled by the atomic size of the NV$^-$ center, its bright fluorescence that allows single centers to be located with nano-resolution and its mechanism of optical spin-polarization/ readout (OSPR) that allows the magnetic resonances of its electron spin to be optically detected (ODMR) \cite{review}.

It is evident that a thorough understanding of the effects of temperature on the NV$^-$ center's electron spin resonances is required to optimize its implementation in quantum technologies for  ambient/ fluctuating environments \cite{fang13}. It is also clear that specific understanding of these effects in nano-diamond NV$^-$ centers (as opposed to bulk diamond) is required for the refined realization of the center's bio-sensing applications. Such specific differences of nano-diamond include variations in thermal properties and generally higher intrinsic strain and abundance of local charge traps. Despite these requirements, only the temperature shift of the zero-field (no strain) ground state electron spin resonance has been examined in detail \cite{toyli12,acostashift1,acostashift2,chenshift,doherty14a}. Consequently, the center's excited state spin resonances represent additional quantum resources for the center's exciting applications that are yet to be precisely examined \cite{fuchs10}.

To rectify this situation, we report comprehensive observations of the effects of temperature on all of the optically detected spin resonances of NV$^-$ centers in nano-diamond. Using models derived from first principles, we describe the temperature variations of each of the zero-field and strain interactions that define the electron spin resonances. In doing so, we provide significant insight into the thermal properties of the resonances and the influence of the nano-diamond environment. Additionally, we demonstrate an alternate all-optical NV$^-$  thermometry technique that exploits the temperature dependence of the center's optical Debye-Waller factor. Our all-optical technique has a projected temperature noise floor of 0.1 K $\mathrm{Hz}^{-1/2}$ for a nano-diamond containing 500 NV$^-$ centers. This noise floor potentially surpasses that of other existing biocompatible nano-thermometry techniques, except for the NV$^-$ ODMR technique \cite{kucsko13,yue12,wang13,wang02}. Indeed, the simplicity and robustness of our all-optical technique may prove advantageous compared to the ODMR technique for particular biological applications where microwave excitation is prohibitive.

\begin{figure}[hbtp]
\mbox{
\subfigure[]{\includegraphics[width=0.25\columnwidth] {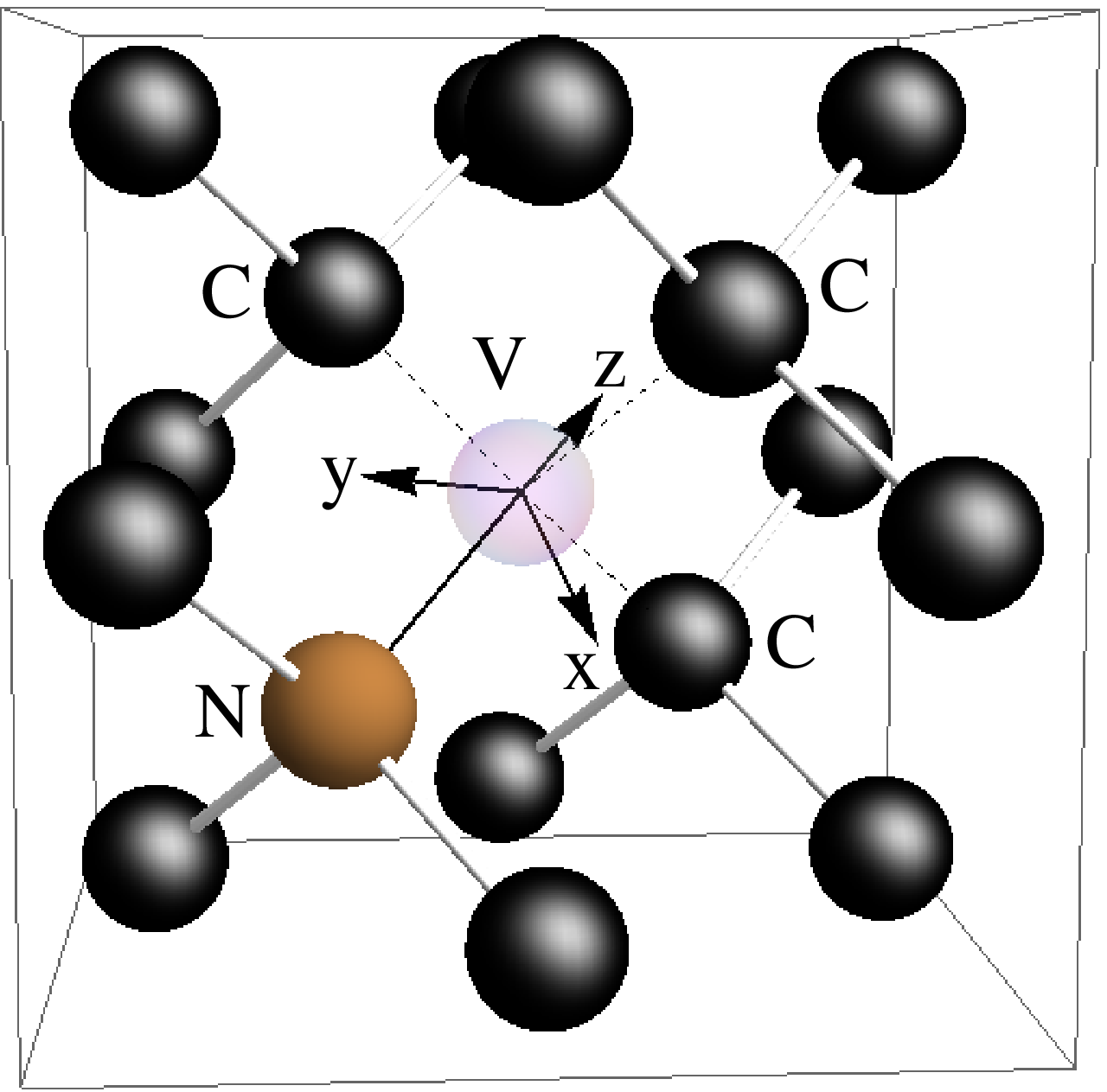}}
\subfigure[]{\includegraphics[width=0.40\columnwidth] {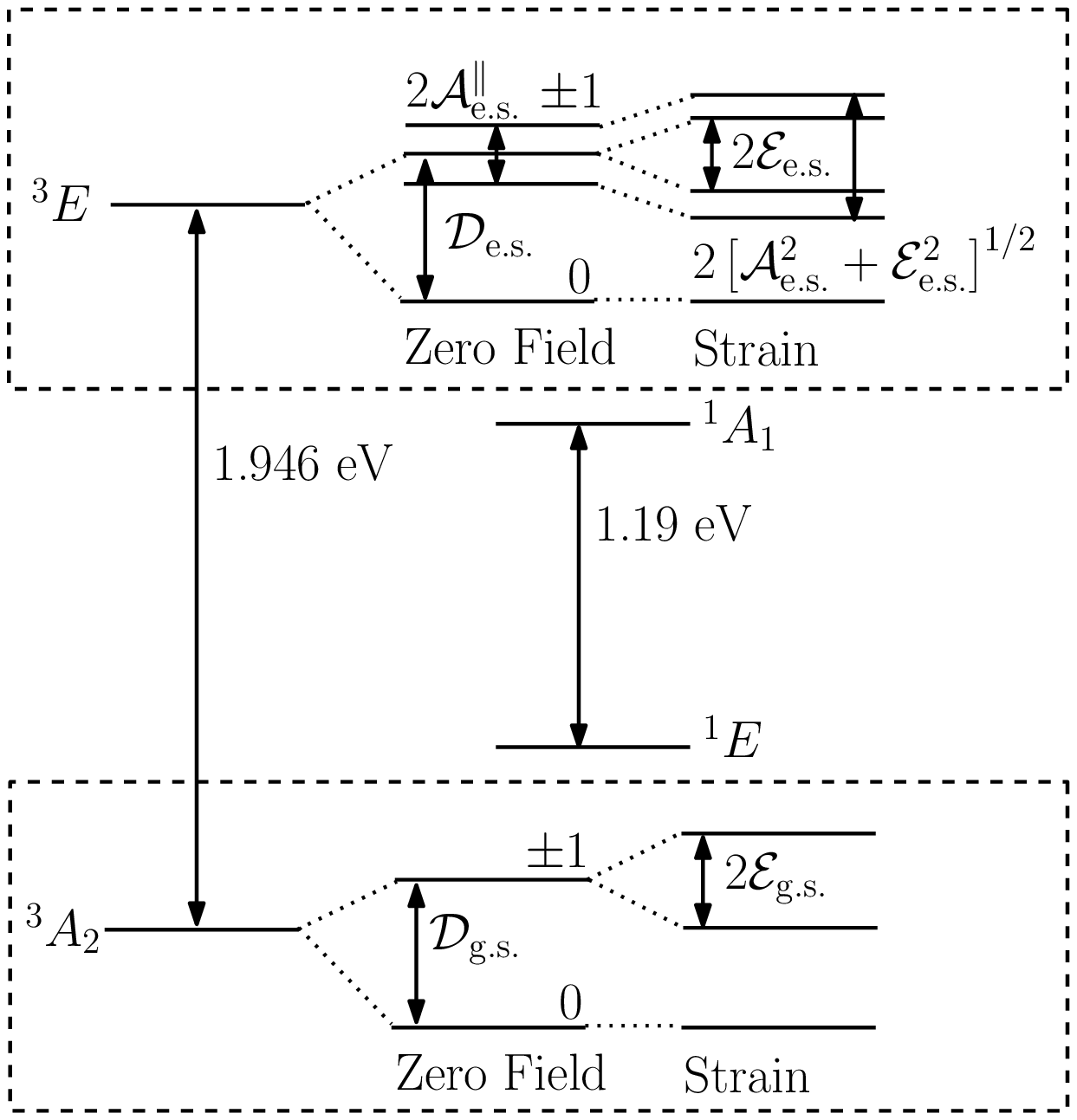}}
\subfigure[]{\includegraphics[width=0.30\columnwidth] {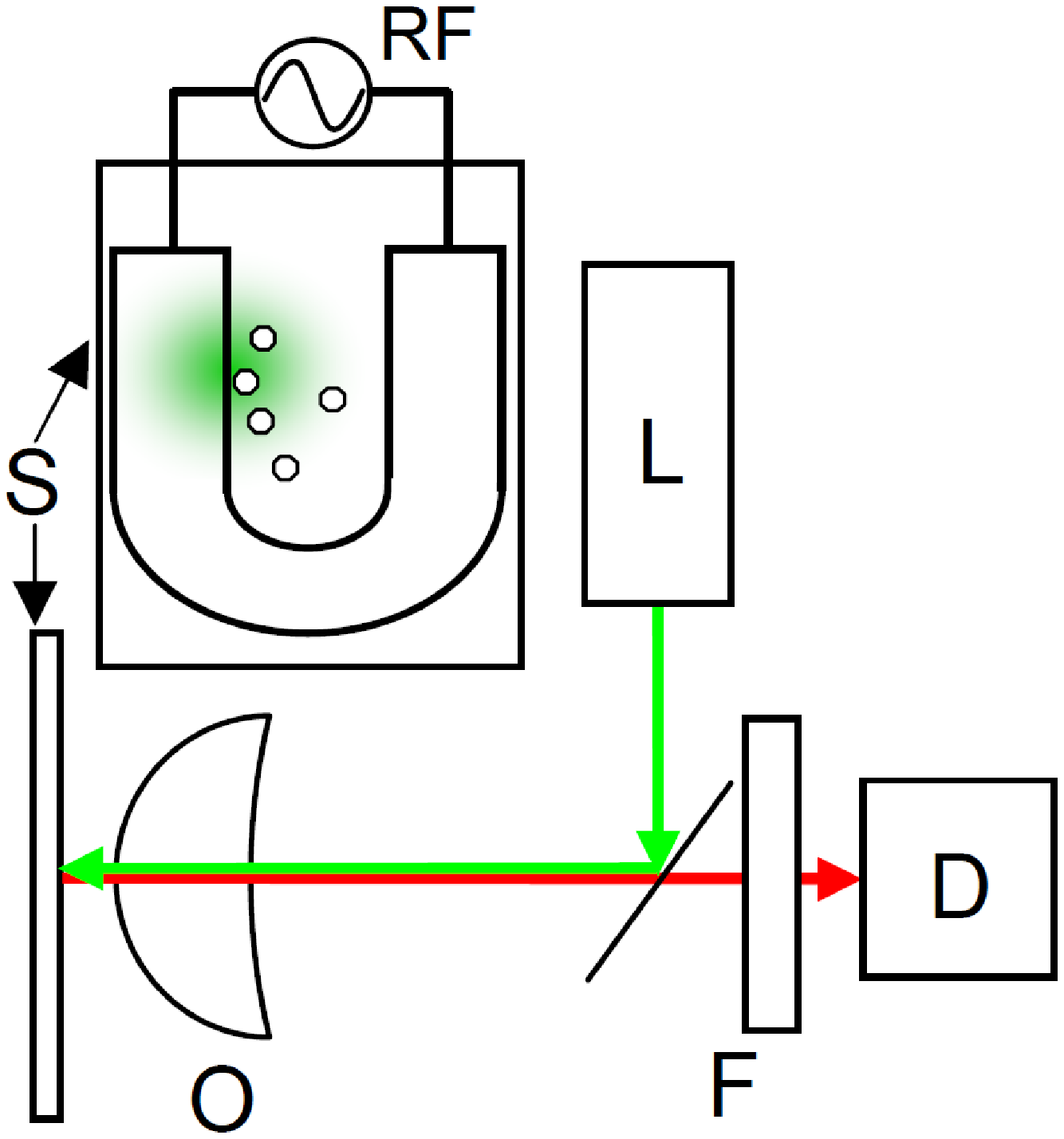}}
}
\caption{(a) Schematic of the NV center depicting the vacancy, the substitutional nitrogen atom, the surrounding carbon atoms and the adopted coordinate system ($z$ axis aligned with the $C_{3v}$ axis of the center and the $x$ axis is contained in one of the center's mirror planes). (b) Schematic of the center's electronic structure, including the low-temperature visible $E_\mathrm{V}\sim$1.946 eV and infrared $E_\mathrm{IR}\sim$1.19 eV ZPL energies. Lower box - The fine structure of the $^3A_2$ level: at zero field with a single splitting ${\cal D}_\mathrm{g.s.}\sim2.87$ GHz (room-temperature); and under symmetry lowering strain, with an additional strain dependent splitting $2{\cal E}_\mathrm{g.s.}$. Upper box - The fine and hyperfine structures of the $^3E$ level observed at room-temperature: at zero field with fine ${\cal D}_\mathrm{e.s.}\sim1.42$ GHz and hyperfine ${\cal A}_\mathrm{e.s.}^\parallel\sim40$ MHz splittings; and under strain, with the additional splitting parameter ${\cal E}_\mathrm{e.s.}$. (c) the experimental setup depicted in two sections: (top) the front projection depicting the laser spot partially covering the U-shaped gold wire connected to a RF generator; (bottom) optical diagram. Labels are defined as: O - microscope objective, D - photodetector, L - laser, F - optical filter and S - sample.
 }
\label{fig:electronicstructure}
\end{figure}

The NV$^-$ center is a $C_{3v}$ point defect in diamond consisting of a substitutional nitrogen atom adjacent to a carbon vacancy that has trapped an additional electron (refer to Figure \ref{fig:electronicstructure}a).  Figure \ref{fig:electronicstructure}b depicts the center's electronic structure, including the low-temperature zero phonon line (ZPL) energies of the visible ($E_\mathrm{V}\sim$1.946 eV)  \cite{davies76} and infrared ($E_\mathrm{IR}\sim$1.19 eV) \cite{rogers08,acosta10b,manson10} transitions. The spin resonances of the $^3A_2$ and $^3E$ levels can be each optically detected at room-temperature as a change in the visible fluorescence intensity and described by a spin-Hamiltonian of the form
\begin{eqnarray}
H & = & {\cal D} \left(S_z^2-\frac{2}{3}\right)+{\cal E}\left(S_y^2-S_x^2\right)+{\cal A}^\parallel S_zI_z+{\cal A}^\perp \left(S_xI_x+S_yI_y\right)
\end{eqnarray}
where $\vec{S}$ and $\vec{I}$ are the dimensionless spin-1 operators of the electron and $^{14}$N nuclear spins, respectively, that are defined with respect to the coordinate system depicted in Figure \ref{fig:electronicstructure}a, and the parameters (${\cal D}$, ${\cal E}$, ${\cal A}^\parallel$, ${\cal A}^\perp$) differ between levels and are identified in the following text.

As depicted in the inset of Figure \ref{fig:electronicstructure}b, the ground $^3A_2$ level exhibits a zero-field fine structure splitting between the $m_s=0$ and $\pm1$ spin sub-levels of ${\cal D}_\mathrm{g.s.}\sim2.87$ GHz (room-temperature) due mainly to electron spin-spin interaction \cite{loubser78}.  Under crystal strain that distorts the $C_{3v}$ symmetry of the center, the $m_s=\pm1$ sub-levels are mixed and their degeneracy is lifted \cite{doherty12}. This strain dependent splitting of the $m_s=\pm1$ sub-levels is $2{\cal E}_\mathrm{g.s.}$. The $^3A_2$ level also exhibits $^{14}$N magnetic hyperfine structure described by the parameters ${\cal A}_\mathrm{g.s.}^\parallel\sim-2.14$ MHz and ${\cal A}_\mathrm{g.s.}^\perp\sim-2.70$ MHz,\cite{felton09} but this minor additional structure does not feature in this work. Nor do the $^{14}$N electric hyperfine interactions. At low temperatures ($<10$ K), the excited $^3E$ level exhibits a non-trivial fine structure arising from a combination of electron spin-orbit, spin-spin and strain interactions \cite{batalov09}. Above $150$ K \cite{batalov09}, the observed fine structure of the $^3E$ level becomes analogous to that of the ground $^3A_2$ level, with a zero-field fine structure splitting between the $m_s=0$ and $\pm1$ spin sub-levels of ${\cal D}_\mathrm{e.s.}\sim1.42$ GHz (room-temperature) and a strain dependent splitting $2{\cal E}_\mathrm{e.s.}$ of the $m_s=\pm1$ sub-levels (see Figure \ref{fig:electronicstructure}) \cite{fuchs08,neumann09}. Above 150 K, the $^3E$ level also exhibits $^{14}$N hyperfine structure analogous to that of the $^3A_2$ level, but with much larger parameters ${\cal A}_\mathrm{g.s.}^\parallel\sim{\cal A}_\mathrm{g.s.}^\perp\sim40$ MHz (assumed isotropic)\cite{steiner10} and is consequently important to this work. The collapse of the observed $^3E$ fine structure above 150 K is thought to be due to phonon mediated orbital averaging, however a precise model of this process is yet to be presented.\cite{rogers09} Consequently, the physical origins of the strain dependent ${\cal E}_\mathrm{e.s.}$ parameter have not been previously established.\cite{review} The $^3E$ hyperfine structure has also not been previously examined theoretically and it has not been observed at low temperatures.\cite{review}

In our experiments, we measured the temperature dependence of the ODMR of NV$^-$ centers in type Ib HPHT nano-diamond crystals (see Figure \ref{fig:electronicstructure}). On average, the nano-diamonds had a diameter of $\sim30$ nm and contained $\sim15$ NV$^-$ centers. We performed ODMR measurements on a total of 10 nano-diamonds. Photoluminescence (PL) spectroscopy was also performed on five of these nano-diamonds and the results of these five nano-diamonds were analysed and found to be consistent. The ODMR results for one representative nano-diamond are presented in the following, whereas PL results for several nano-diamonds are presented as specified. The nano-diamonds were spin coated on a fused quartz substrate. The spin resonances were driven by microwaves carried by a U-shaped, 4 cm long and $200\times0.5$ $\mu$m in cross-section band of deposited gold on the substrate. The investigated crystals were located at distances between 3 and 30 $\mu$m from the edge of the gold band. The spin resonances were optically detected using 532 nm laser excitation and fluorescence collection via an epifluorescence design. ODMR spectra were measured using continuous wave (c.w.) optical and microwave excitation. As the microwave field was weak and the crystals had good thermal contact with the substrate, the crystals were primarily heated (up to 600 K) by the laser light illumination of the gold band. See supplementary information for further experimental details.

\begin{figure}[hbtp]
\mbox{
\subfigure[]{\includegraphics[width=0.33\columnwidth] {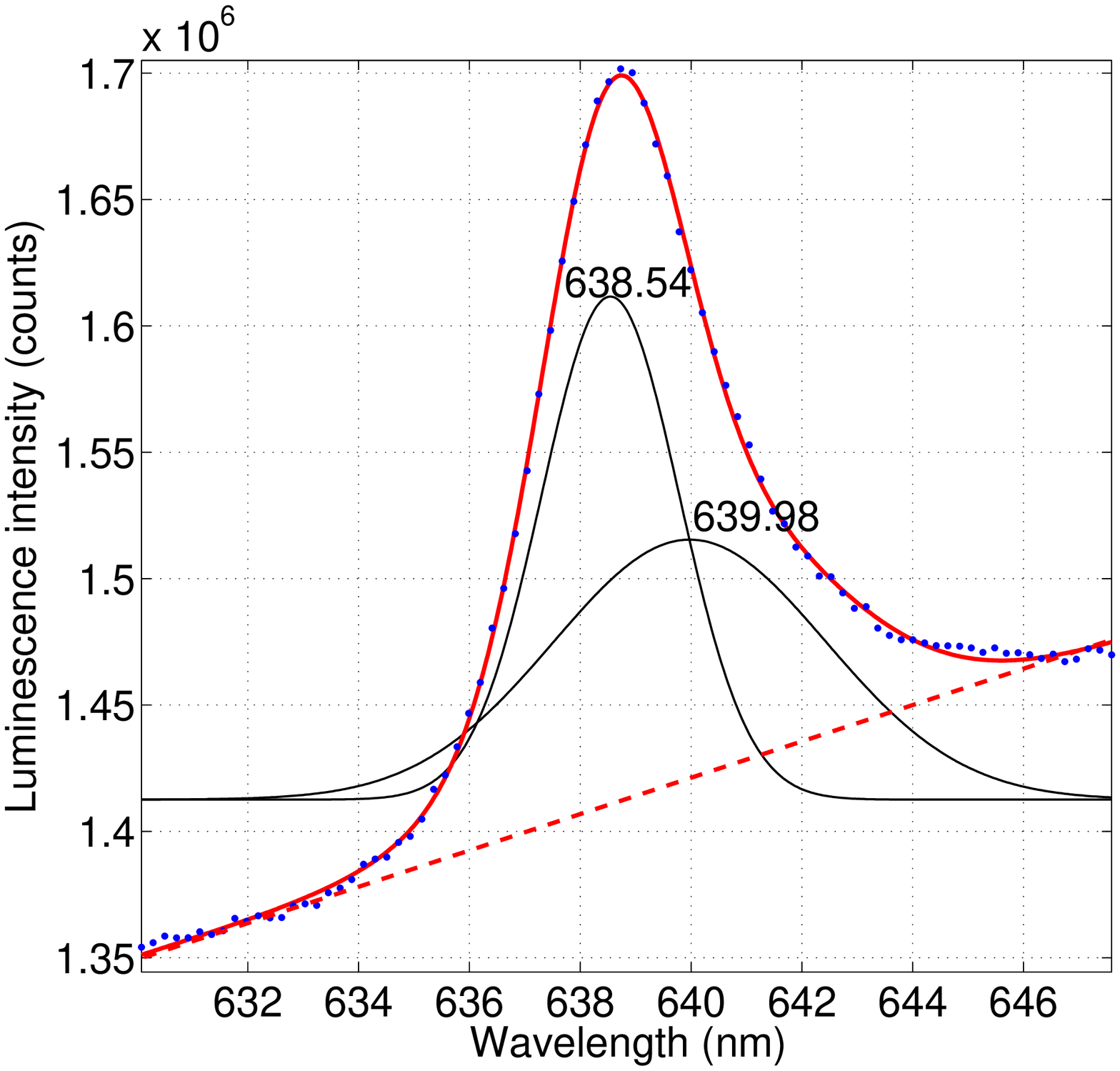}}
\subfigure[]{\includegraphics[width=0.33\columnwidth] {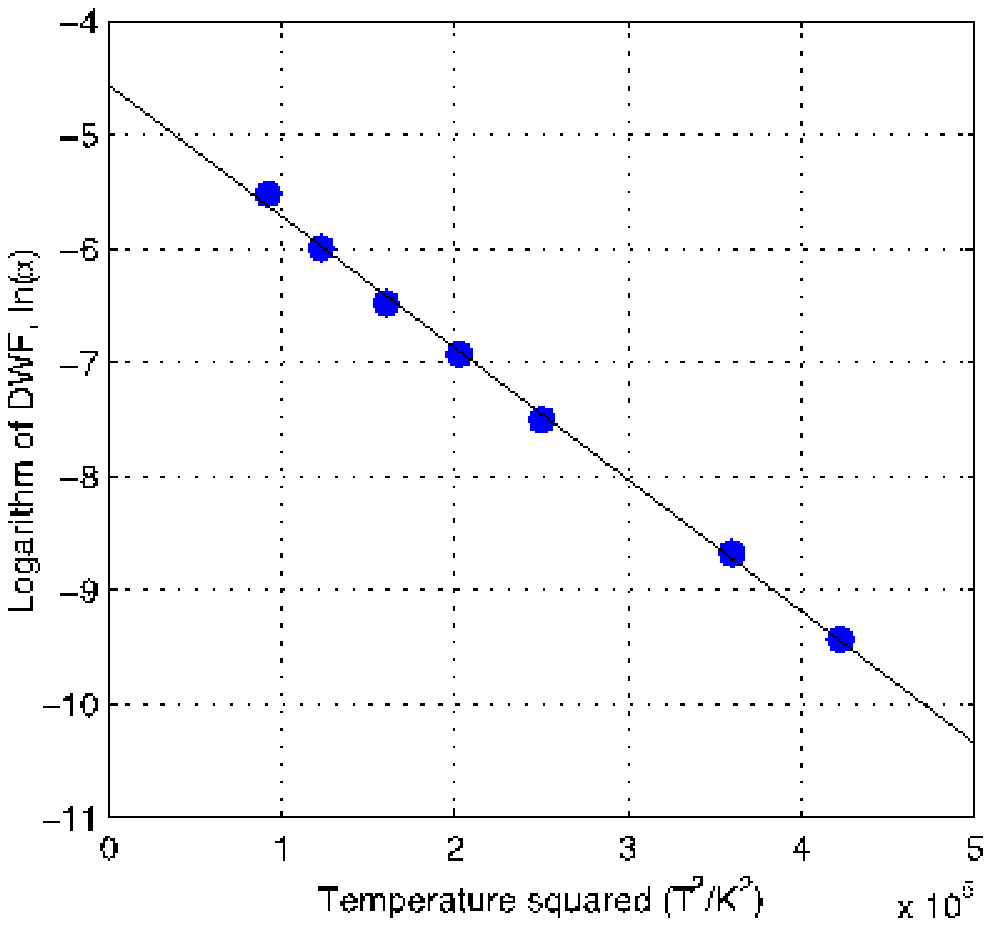}}
\subfigure[]{\includegraphics[width=0.33\columnwidth] {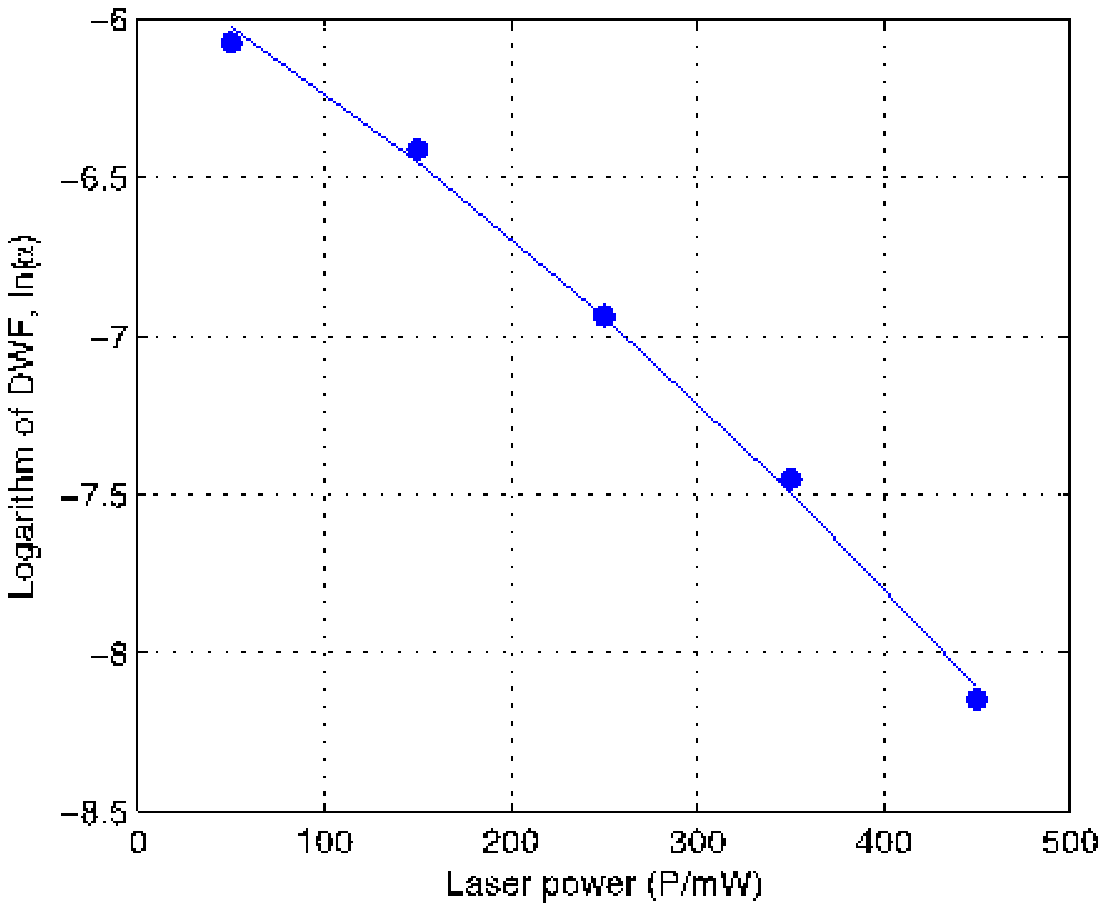}}
}
\caption{Example NV$^-$ visible ZPL spectrum (a) at room-temperature and the DWF as a function of (b) oven temperature $T$ and (c) laser power $P_\mathrm{las.}$. (a) demonstrates the fit (red solid) of the ZPL (blue points) using two strain-split lorentzian lineshapes (black solid) and a linear function (red dashed) to account for the contribution of the phonon sideband underneath the ZPL. The equation of the straight line in (b) is $\log \mathrm{DWF} = -S(1+\frac{2}{3}\pi^2T^2/T_D^2)$, where $S=4.57(7)$ and $T_D=1614(23)$ K are derived from a least-squares fit to the data points (circles). The dependence of the DWF on the laser power in (c) is described by the substitution $T=T_0+bP_\mathrm{las.}$, where $T_0=294$ K is the room-temperature and a least-squares fit of the data points (circles) yields $S=4.79(6)$ and $b=0.51(3)$ K/ mW. In (b) and (c), error bars are smaller than the point size (see supplementary information).}
\label{fig:DWF_thermometry}
\end{figure}

Our all-optical nano-thermometry technique uses the Debye-Waller factor (DWF) of the NV$^-$ visible transition. The DWF is the ratio of the area under the ZPL and the total emission band (see Figure \ref{fig:DWF_thermometry}). The technique was calibrated by measuring the DWF of nano-diamond NV$^-$ centers inside an oven operating under stabilized temperature conditions (see Figure \ref{fig:DWF_thermometry}). For $T\ll T_D$, the temperature dependence of the DWF is well described by the simple Debye model of the phonon continuum \cite{brand81,richards71,fitchen63}
\begin{eqnarray}
\mathrm{DWF} & = & e^{-S(1+\frac{2}{3}\pi^2 T^2/T_D^2 )}
\end{eqnarray}
where $T_D$ is the Debye temperature of the host crystal ($T_D\sim2220$ K for bulk diamond) and $S$ is a parameter defining the electron-phonon coupling strength. For the nano-diamonds in the oven, the observed value of $T_D$ was $\sim1614(23)$ K, which is 1.37 times smaller than the value for bulk diamond derived from its thermodynamic properties at low temperatures. Such a mismatch is typical and is well established in other materials \cite{fitchen63}. Given the observed value of $T_D$, the temperature of the nano-diamonds subjected to the laser-assisted heating was found to follow a linear function $T=T_0+bP_\mathrm{las.}$ of laser power $P_\mathrm{las.}$, where $T_0\sim 294$ K is the room-temperature and the proportionality constant $b$ was found to be 0.51(3) K/ mW for the crystal whose data is shown in  Figure \ref{fig:DWF_thermometry}c. The electron-phonon coupling parameter $S$ was determined separately for the nano-diamonds in the oven and those used for the ODMR experiments.

Given the collection of $N$ photons emitted by NV$^-$ centers in a nano-diamond, the smallest temperature change that can be detected by the optical DWF thermometry technique is defined by the shot-noise uncertainty in the measurement and the temperature dependence of the DWF
\begin{equation}
\delta T_\mathrm{min} = \frac{\mathrm{DWF}+\sqrt{\mathrm{DWF}}}{\sqrt{N}}\left(\frac{d \mathrm{DWF}}{d T}\right)^{-1}\approx \frac{\sqrt{\mathrm{DWF}}}{\sqrt{N}}\left(\frac{d \mathrm{DWF}}{d T}\right)^{-1}.
\end{equation}
The presence of background fluorescence raises the uncertainty by a factor of $\sqrt{1+3r}$, where $r$ is the ratio of the approximately uniform background intensity under the NV$^-$ ZPL and the ZPL peak intensity.\cite{donley01} The value of $N$ is related to the measurement time $\tau$, the optical collection efficiency $\mu$, the single NV$^-$ photon emission rate $\gamma$ and the number of NV$^-$ centers $n$ via $N=n\mu\gamma\tau$. Consequently, the temperature noise floor $\eta_T$ of the DWF thermometry technique is
\begin{equation}
\eta_T = \delta T_\mathrm{min}\sqrt{\tau} = \sqrt{(1+3r)}C_\mathrm{ZPL}^{-1/2}\Phi,
\end{equation}
where $C_\mathrm{ZPL}=n\mu\gamma \mathrm{DWF}$ is the ZPL emission rate and $\Phi=\mathrm{DWF} (d\mathrm{DWF}/dT)^{-1}$. At room-temperature, $\Phi$ exhibited little variation over our nano-diamond sample and had an average value of $\Phi=154(10)$ K. Using the typical optical lifetime of NV$^-$ centers in nano-diamond\cite{beveratos02} and the details of the optical setup,\cite{chapman11} we can estimate $\gamma\sim40$ MHz, $\mu\sim0.021$ and DWF$\sim0.005$. The estimated noise floor of a single NV$^-$ center at room-temperature and in the absence of background fluorescence is thus $\eta_T \sim 2.3$ K Hz$^{-1/2}$. If $n\sim500$ as in the recent bio-thermometry demonstration,\cite{kucsko13} the the noise floor becomes $\eta_T \sim 0.1$ K Hz$^{-1/2}$.

\begin{figure}[hbtp]
\mbox{
\subfigure[]{\includegraphics[width=0.4\columnwidth] {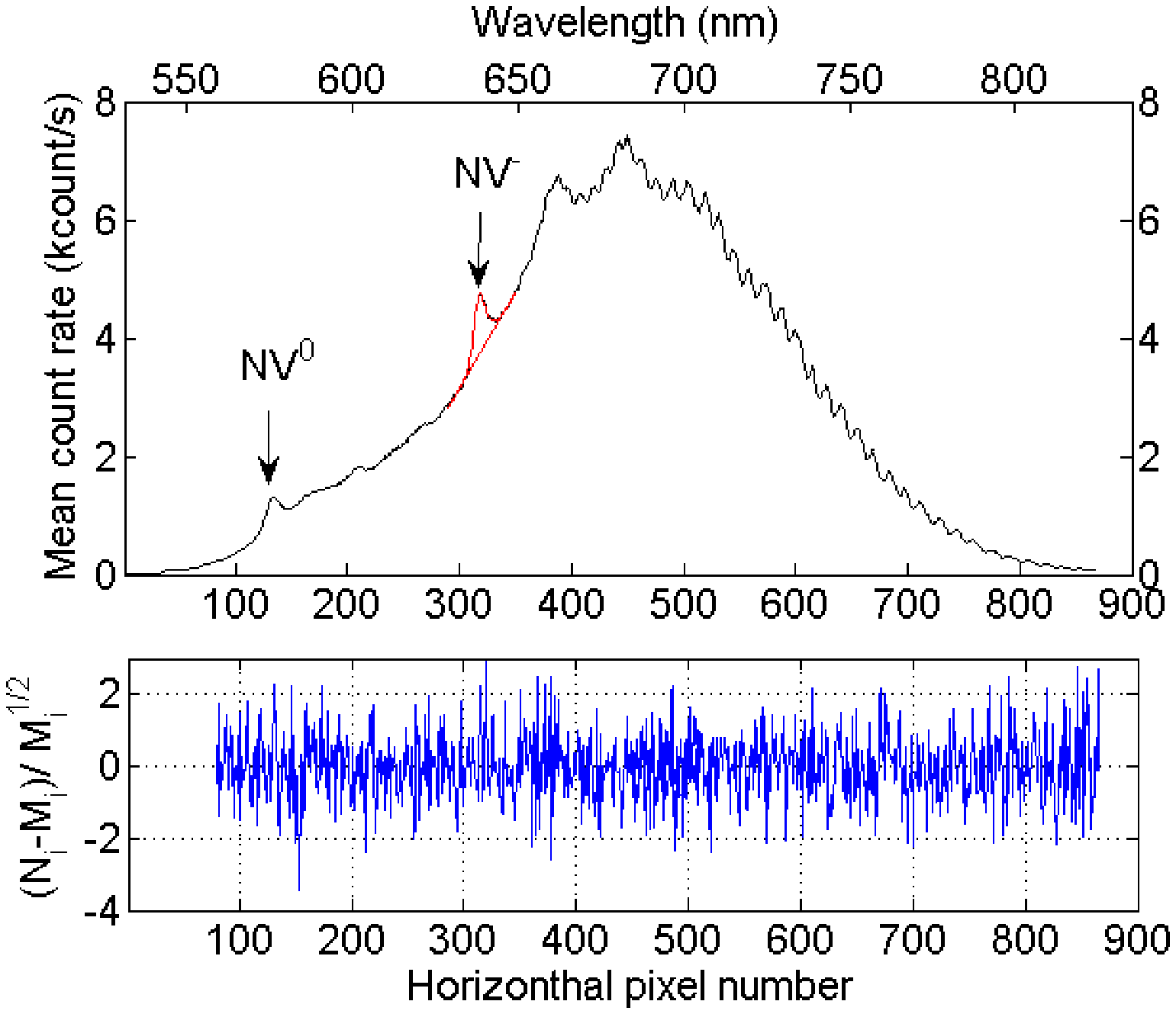}}
\subfigure[]{\includegraphics[width=0.39\columnwidth] {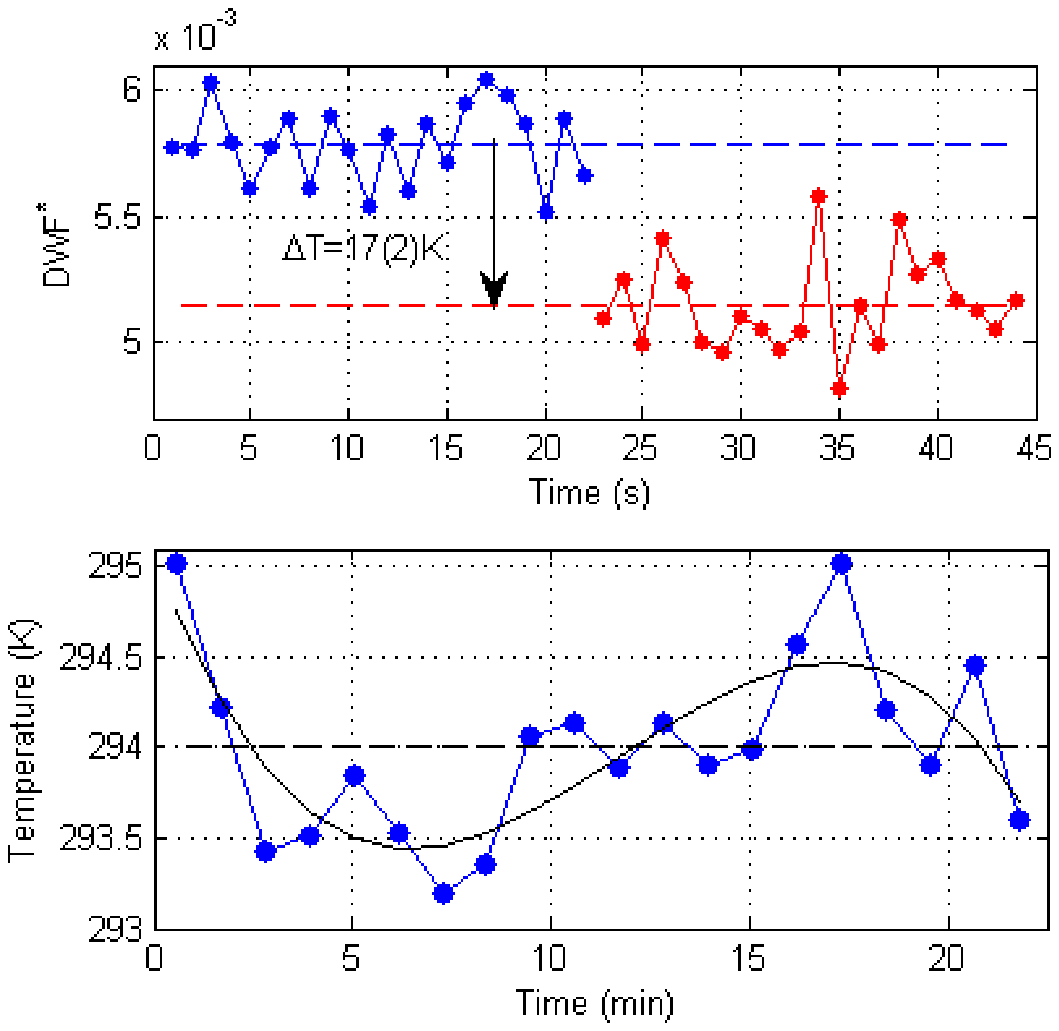}}
}
\caption{(a) Upper panel: PL spectrum collected over one second at room-temperature (upper panel) of a nano-diamond containing NV$^-$ and NV$^0$ centers. The ZPLs of the NV charge states are as denoted and the NV$^-$ ZPL area is outlined in red. Lower panel: the standard deviations of the number of counts per spectrum pixel $N_i$ from its mean value $M_i$ determined by a spectrum collected over a much longer time (50 s). The equality of the standard deviation of $(N_i-M_i)/\sqrt{M_i}$ to 1 demonstrates the Poissonian noise of the spectrum. (b) Upper panel: Time series of DWF$^\ast$ measurements that were each obtained after one second of averaging. DWF$^\ast$ is distinguished from DWF and is defined as the NV$^-$ ZPL area divided by the area of the total fluorescence spectrum (including the background NV$^0$ fluorescence). Note that $\mathrm{DWF}^\ast(d\mathrm{DWF}^\ast/dT)^{-1}=\mathrm{DWF}(d\mathrm{DWF}/dT)^{-1}=\Phi$. A temperature step of 17(2) K above room-temperature (as measured via DWF technique), achieved via laser heating, is evident at $t=23$ s. Lower panel: Time series of DWF temperature measurements (points) over 23 minutes under ambient conditions with a time-averaged temperature (dashed line) of 294 K and 0.5 K fluctuations. The solid cubic-polynomial line indicates possible long-time trend. Each measurement was averaged over 50 s.}
\label{fig:DWFdemonstration}
\end{figure}

To test our estimate of the noise floor of the DWF thermometry technique, we characterized the noise in our DWF measurements and performed a series of temperature measurements. Figure \ref{fig:DWFdemonstration}a demonstrates the Poissonian fluctuations of a fluorescence spectrum of a nano-diamond. This spectrum was collected over one second and contains fluorescence from NV$^-$ centers as well some neutral NV (NV$^0$) centers present in this particular nano-diamond. The upper panel of figure \ref{fig:DWFdemonstration}b depicts a sequence of DWF measurements, including a step caused by a 17 K temperature increase via laser heating. Each measurement was averaged over one second. The observed standard deviation of the DWF measurements corresponds to a temperature uncertainty of 4.0(4) K, which directly implies a temperature noise floor of 4.0(4) K Hz$^{-1/2}$. The lower panel of figure \ref{fig:DWFdemonstration}b depicts a series of temperature measurements using a second nano-diamond at room-temperature over a period of 23 minutes. This nano-diamond contained more NV$^-$ and NV$^0$ centers and each measurement was averaged over 50 s. The standard deviation of the measurements from the mean value was 0.50(5) K, which corresponds to a noise floor of 3.5(3) K Hz$^{-1/2}$. The deviation from a cubic polynomial trend of the room temperature is even smaller, 0.3 K.  The noise floors achieved using these two nano-diamonds are in agreement with (4). This can be confirmed using the experimentally observed values of $C_\mathrm{ZPL}\sim$ 14 kHz and 42 kHz and $r\sim$ 3.7 and 10.5 for these nano-diamonds, respectively. As seen in figure \ref{fig:DWFdemonstration}a, the non-zero NV$^0$ fluorescence makes the main contribution to the background. The increased noise floor arising from the NV$^0$ background can be reduced in several ways. Since the NV$^0$ center ZPL occurs at 575 nm, NV$^0$ fluorescence can be avoided if optical excitation is instead performed at wavelengths in the range $575-637$ nm. Alternatively, nano-diamonds with relatively few NV$^0$ centers could be purposefully fabricated or post-selected. Hence, NV$^0$ background fluorescence does not present a fundamental barrier to achieving the projected noise floor $\eta_T \sim 0.1$ K Hz$^{-1/2}$ for a nano-diamond containing $\sim$500 NV$^-$ centers.

This projected sub-Kelvin noise floor of the DWF thermometry technique is comparable to the demonstrated 130 mK Hz$^{-1/2}$ noise floor of the ODMR technique in nano-diamond\cite{neumann13} and it potentially surpasses the noise floors of other biocompatible nano-thermometry techniques with sensor sizes less than 100 nm\cite{kucsko13,yue12,wang13,wang02}. The notable competitor being CdX (X= S, Se or Te) quantum dots, whose noise floors have not been clearly determined, but based upon some demonstrations, could rival both NV$^-$ nano-thermometry techniques \cite{wang13,wang02}. The projected noise floor of the ODMR technique for ideal nano-diamonds containing 500 NV$^-$ centers is $\sim$1 mK Hz$^{-1/2}$ \cite{kucsko13}. Whilst this is much lower than the noise floor of the DWF technique, the fabrication of such ideal nano-diamonds, with negligible strain inhomogeneity and sufficient purity to support NV$^-$ spin coherence times of $\sim 1$ ms, is yet to be achieved. Comparing further with the ODMR technique, the DWF technique is a much simpler and more flexible all-optical technique that does not require microwave control, and thus may be better suited to particular biological applications. Indeed, the DWF technique can be immediately implemented using a commercial spectral-resolving imaging system, such as a Raman microscope.  Like the ODMR technique, the DWF technique has the potential to be extended to \textit{in vivo} applications if a different method of fluorescence spectroscopy was employed, such as two-photon excitation spectroscopy \cite{helmchen05,wee07}. Spatial resolution may also be further improved by all-optical far-field sub-diffraction imaging techniques \cite{rittweger09a,rittweger09b}. Thus, there is clear motivation for further investigation of the DWF thermometry technique as an alternate/complementary technique to the ODMR technique and other current nano-thermometry techniques.

There has been several studies of the effects of pressure and temperature on the visible ZPL, the infrared ZPL and the zero-field $^3A_2$ spin resonance ${\cal D}_\mathrm{g.s.}$ of centers in bulk diamond.\cite{toyli12,acostashift1,acostashift2,chenshift,doherty14a,davies74} Acosta et al have also studied the temperature dependence of the $^3A_2$ strain splitting ${\cal E}_\mathrm{g.s.}$ in bulk diamond NV$^-$ ensembles, but since it was much smaller than ${\cal D}_\mathrm{g.s.}$, did not observe a variation above experimental uncertainty over the range 5-400 K.\cite{acostashift1,acostashift2} The pressure and temperature dependence of the $^3E$ splittings $D_\mathrm{e.s.}$ and ${\cal E}_\mathrm{e.s.}$ have not been previously studied. Recently, Doherty et al \cite{doherty14a} demonstrated that the model original developed by Davies \cite{davies74} to explain the temperature shift of the visible ZPL energy, successfully describes the temperature shifts of the infrared ZPL energy and ${\cal D}_\mathrm{g.s.}$. In Davies' model, there are two origins of the temperature shifts of the spin resonances: (1) the spin energies are perturbed by the strain of thermal expansion, and (2) the vibrational frequencies associated with different spin states differ. The later is a consequence of quadratic electron-phonon interactions.\cite{doherty14a} Since thermal expansion and hydrostatic pressure are intimately related, the contribution of thermal expansion to the temperature shifts may be alternatively expressed in terms of the hydrostatic pressure shifts and the pressure of thermal expansion in diamond \cite{davies74}. Thus, the temperature shift of the $^3A_2$ zero-field resonance $\Delta {\cal D}_\mathrm{g.s.}(T)= \Delta {\cal D}_\mathrm{g.s.}^\mathrm{ex}(T)+\Delta {\cal D}_\mathrm{g.s.}^\mathrm{e-p}(T)$ is the sum of thermal expansion $\Delta {\cal D}_\mathrm{g.s.}^\mathrm{ex}(T)$ and quadratic electron-phonon $\Delta {\cal D}_\mathrm{g.s.}^\mathrm{e-p}(T)$ contributions given by
\begin{eqnarray}
\Delta {\cal D}_\mathrm{g.s.}^\mathrm{ex}(T) & = & \Gamma_\mathrm{g.s.} P(T) \nonumber \\
\Delta {\cal D}_\mathrm{g.s.}^\mathrm{e-p}(T) & = & \int_0^\Omega n(\omega,T)\delta_\mathrm{g.s.}(\omega)\rho(\omega)d\omega,
\end{eqnarray}
where $\Gamma_\mathrm{g.s.}=14.58(6)$ MHz/GPa\cite{doherty14} is the hydrostatic pressure shift of ${\cal D}_\mathrm{g.s.}$,  $P(T) = B\int_0^T e(t)dt$ is the pressure of thermal expansion, $B = 442$ GPa is the bulk modulus of diamond, $e(T)$ is the diamond volume expansion coefficient, $n(\omega,T) = (e^{h\omega/k_B T}-1)^{-1}$ is the Bose-Einstein distribution of vibrational occupations, $h$ is Planck's constant, $k_B$ is Boltzmann's constant, $\Omega$ is the highest vibrational frequency of diamond, $\rho(\omega)$ is the vibrational density of states, and $\delta_\mathrm{g.s.}(\omega)$ is  related to the differences in the vibrational frequencies of the $^3A_2$ electron spin states.\cite{doherty14a} Given power series expansions of $e(T)$ and $\delta_\mathrm{g.s.}(\omega)\rho(\omega)$, the temperature shift $\Delta {\cal D}_\mathrm{g.s.}(T)$ may be approximated by a polynomial in temperature $T$. \cite{doherty14a} Doherty et al's application of Davies' model is extended here to ${\cal E}_\mathrm{g.s.}$ and ${\cal D}_\mathrm{e.s.}$ by deriving analogous expressions for $\Delta {\cal D}_\mathrm{e.s.}(T)$ and $\Delta {\cal E}_\mathrm{g.s.}(T)$, where the corresponding $\Gamma$ and $\delta(\omega)$ parameters differ from $\Delta {\cal D}_\mathrm{g.s.}(T)$. Explicit expressions for each of the $\Gamma$ and $\delta(\omega)$ parameters can be obtained from first-principles via the application of the molecular orbital model of the center (see supplementary information) \cite{doherty11,maze11}.

A model of the collapse of the $^3E$ fine structure above 150 K and the temperature dependence of ${\cal E}_\mathrm{e.s.}(T)$ can be derived from the low temperature $^3E$ fine structure in the condition where spin-conserving electron-phonon transitions within the $^3E$ level occur much faster than the frequencies of the $^3E$ fine structure interactions and the optical decay rate (see supplementary information). Observations of the temperature dependence of the $^3E$ ODMR intensity \cite{batalov09} and the visible ZPL width \cite{fu09} imply that this condition is satisfied at temperatures $> 150$ K. Recognising that the $^3E$ fine structure can be described by the interaction of orbital and spin sub-systems, this condition implies that thermal equilibrium of the $^3E$ orbital states is achieved within the lifetime of the $^3E$ level and that no coherences exist between the orbital and spin sub-systems.\cite{slitcher} Hence, the orbital and spin sub-systems may be decoupled, which leads to the expression
\begin{eqnarray}
{\cal E}_\mathrm{e.s.}(T) & = & D_\mathrm{e.s.}^\perp {\cal R}(T)
\end{eqnarray}
where $D_\mathrm{e.s.}^\perp\sim0.775$ GHz is the low temperature transverse electron spin-spin interaction of the $^3E$ level \cite{batalov09} and ${\cal R}(T) = (e^{h\xi_\perp/k_B T}-1)/(e^{h\xi_\perp/k_B T}+1)$ is a strain-temperature reduction factor derived from the Boltzmann distribution of the orbital sub-system and $\xi_\perp$ is the strain splitting of the $^3E$ fine structure. The above expression clearly demonstrates that the physical origins of ${\cal E}_\mathrm{e.s.}$ is a product of transverse spin-spin interactions $D_\mathrm{e.s.}^\perp$ and a factor ${\cal R}$ that explicitly depends on the ratio of the $^3E$ strain interaction and temperature. In addition to the explicit factor ${\cal R}$, ${\cal E}_\mathrm{e.s.}$ is intrinsically dependent on strain and temperature via their effect on the spin-spin interaction $D_\mathrm{e.s.}^\perp$. This intrinsic dependence will be analogous to the previously described strain and temperature variations of the other parameters ${\cal D}_\mathrm{g.s.}$, ${\cal D}_\mathrm{e.s.}$ and ${\cal E}_\mathrm{g.s.}$.

\begin{figure}[t]
\begin{center}
\mbox{
\subfigure[]{\includegraphics[width=0.33\columnwidth] {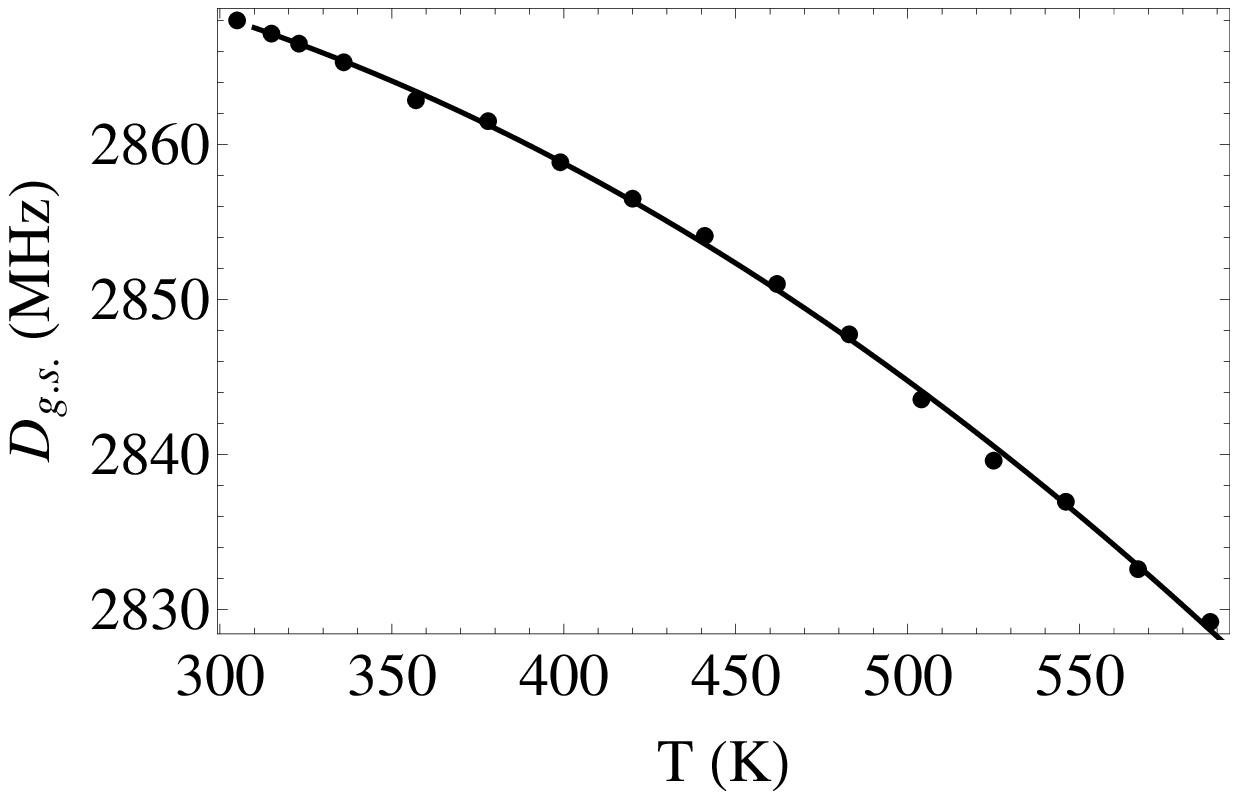}}
\subfigure[]{\includegraphics[width=0.33\columnwidth] {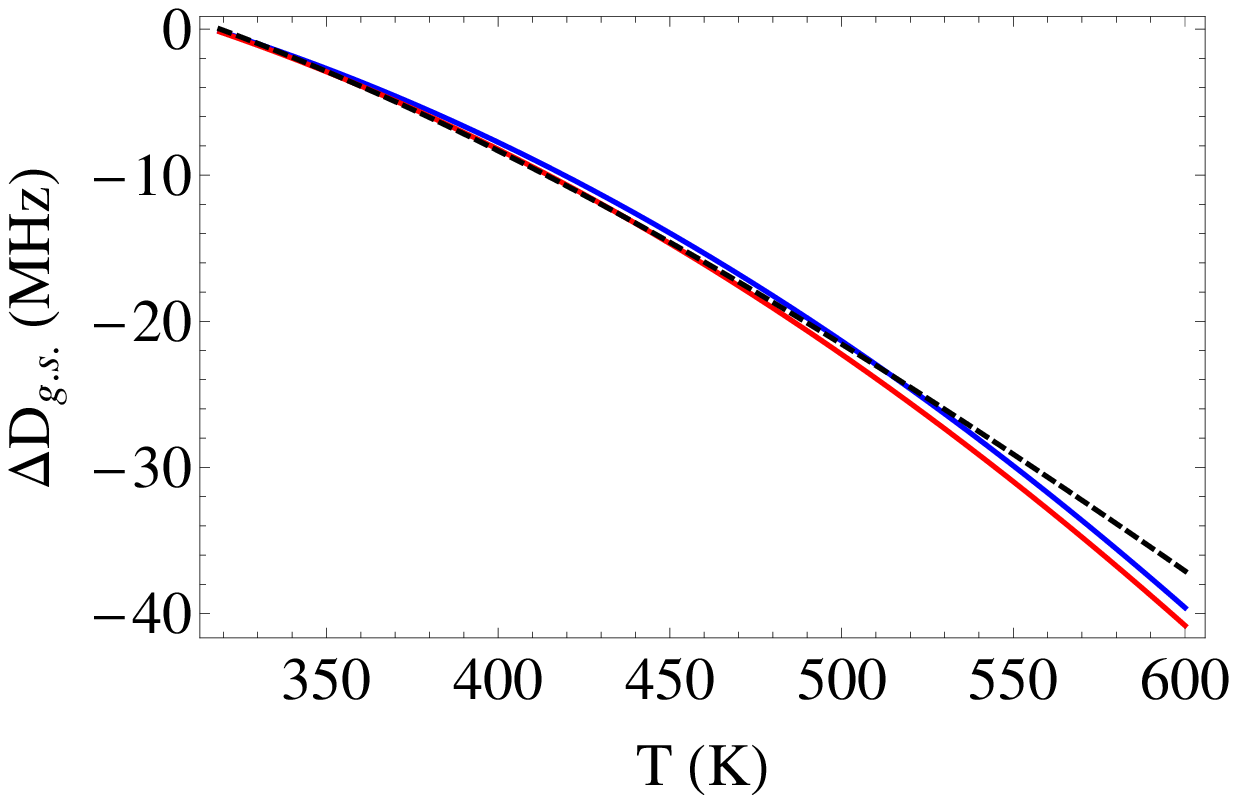}}
}
\caption{(a) The observed temperature variation of ${\cal D}_\mathrm{g.s.}$ and the fit of a quadratic polynomial ${\cal D}_\mathrm{g.s.} = a_\mathrm{g.s.}+b_\mathrm{g.s.} T+c_\mathrm{g.s.} T^2$, where $a_\mathrm{g.s.}=2870(3)$ MHz, $b_\mathrm{g.s.}=6(1)\times10^{-2}$ MHz/K and $c_\mathrm{g.s.}=-2.3(2)\times10^{-4}$ MHz/K$^2$. (b) Comparison of the temperature shift $\Delta {\cal D}_\mathrm{g.s.}(T)$ observed here in two different nano-diamonds (blue and red solid) and the previous observation in bulk diamond (black dashed) \cite{toyli12}. Error bars in (a) are smaller than the point size (see supplementary information).}
\label{fig:groundstateshift}
\end{center}
\end{figure}

The observed temperature variation of the ground state $^3A_2$ spin parameter ${\cal D}_\mathrm{g.s.}$ is depicted in Figure \ref{fig:groundstateshift}a. Referring to Figure \ref{fig:groundstateshift}b, it can be seen that over the majority of the temperature range, the temperature shift $\Delta {\cal D}_\mathrm{g.s.}(T)= {\cal D}_\mathrm{g.s.}(T)-{\cal D}_\mathrm{g.s.}(294 \mathrm{K})$ observed here in nano-diamond is very similar to the previous observation in bulk diamond \cite{toyli12}, thus implying that there is little variance of $\Delta {\cal D}_\mathrm{g.s.}(T)$ in nano-diamond. This conclusion is important to the large-scale implementation of NV$^-$ nano-thermometers based upon $\Delta {\cal D}_\mathrm{g.s.}(T)$ since it suggests that there will be little variation between nano-thermometers.\cite{doherty14a} Furthermore, as temperature in our nano-diamond $\Delta {\cal D}_\mathrm{g.s.}(T)$ measurements was calibrated using the optical DWF thermometry technique, the similarity of our observations to those in bulk diamond further validate the optical DWF technique.

As experienced by Acosta et al in the bulk diamond measurements over temperature 5-400 K,\cite{acostashift1,acostashift2} we did not conclusively observe a temperature variation of ${\cal E}_\mathrm{g.s.}$ up to 600 K. The difficulty in observing the temperature shift of ${\cal E}_\mathrm{g.s.}$ arises from its much smaller magnitude ($\sim 10$ MHz here), which based upon the $0.01$ fractional change of ${\cal D}_\mathrm{g.s.}$ over the range 300-600 K, implies a variation of ${\cal E}_\mathrm{g.s.}$ at the limit of detection ($\sim 0.1$ MHz here). The first-principles model of ${\cal E}_\mathrm{g.s.}$ identifies that the temperature variation of ${\cal E}_\mathrm{g.s.}$ is potentially much more complicated than ${\cal D}_\mathrm{g.s.}$ due to a mixture of strain and transverse spin-spin factors (see supplementary information for discussion). The approximate temperature independence of ${\cal E}_\mathrm{g.s.}$ in nano-diamond is, however, an important observation for ODMR thermometry techniques that utilize a single ground state spin resonance (with frequency ${\cal D}_\mathrm{g.s.}\pm{\cal E}_\mathrm{g.s.}$ in the absence of a magnetic field), since the temperature variation of the spin resonance is restricted to just that of ${\cal D}_\mathrm{g.s.}$.

\begin{figure}[hbtp]
\begin{center}
\mbox{
\subfigure[]{\includegraphics[width=0.33\columnwidth] {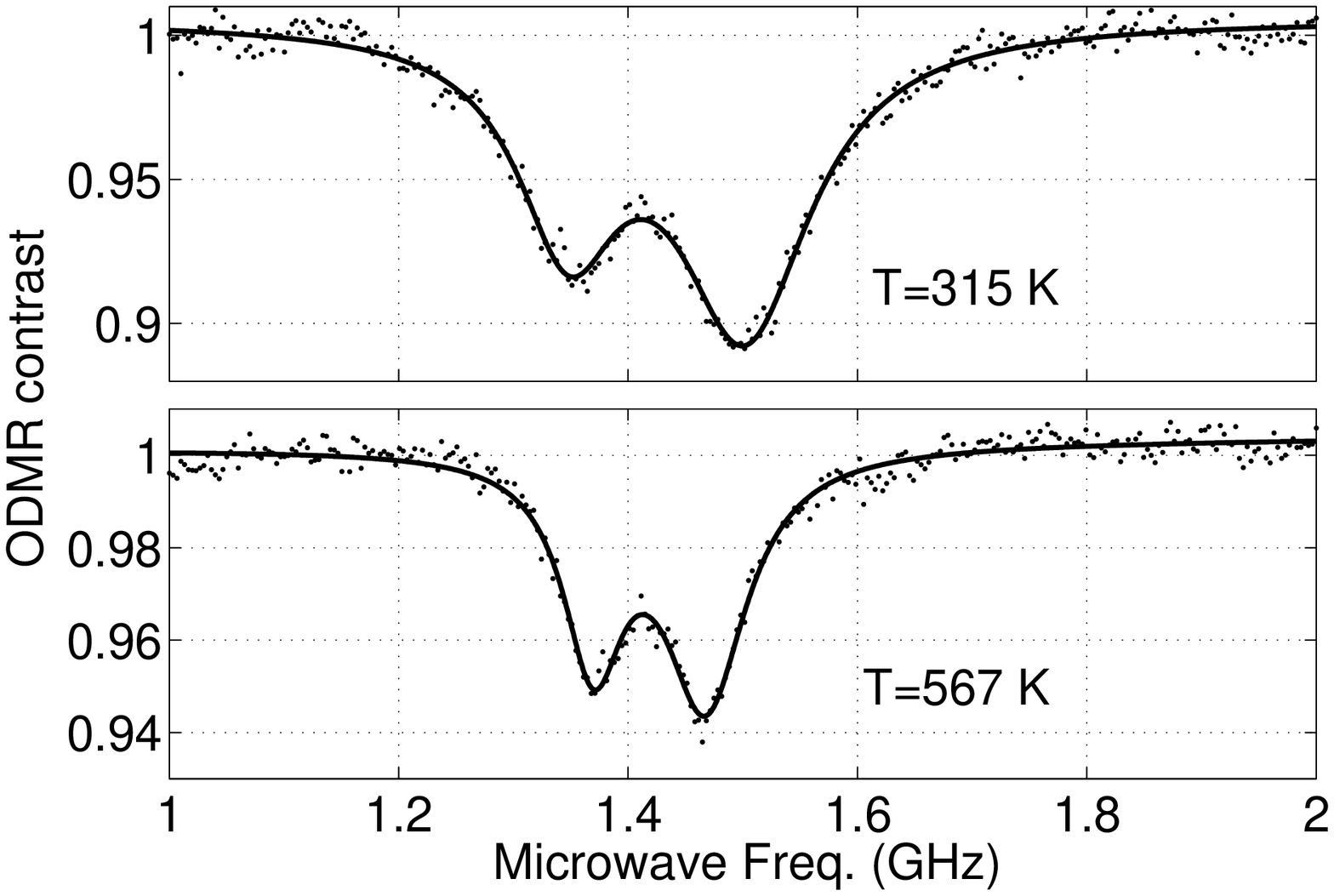}}
\subfigure[]{\includegraphics[width=0.33\columnwidth] {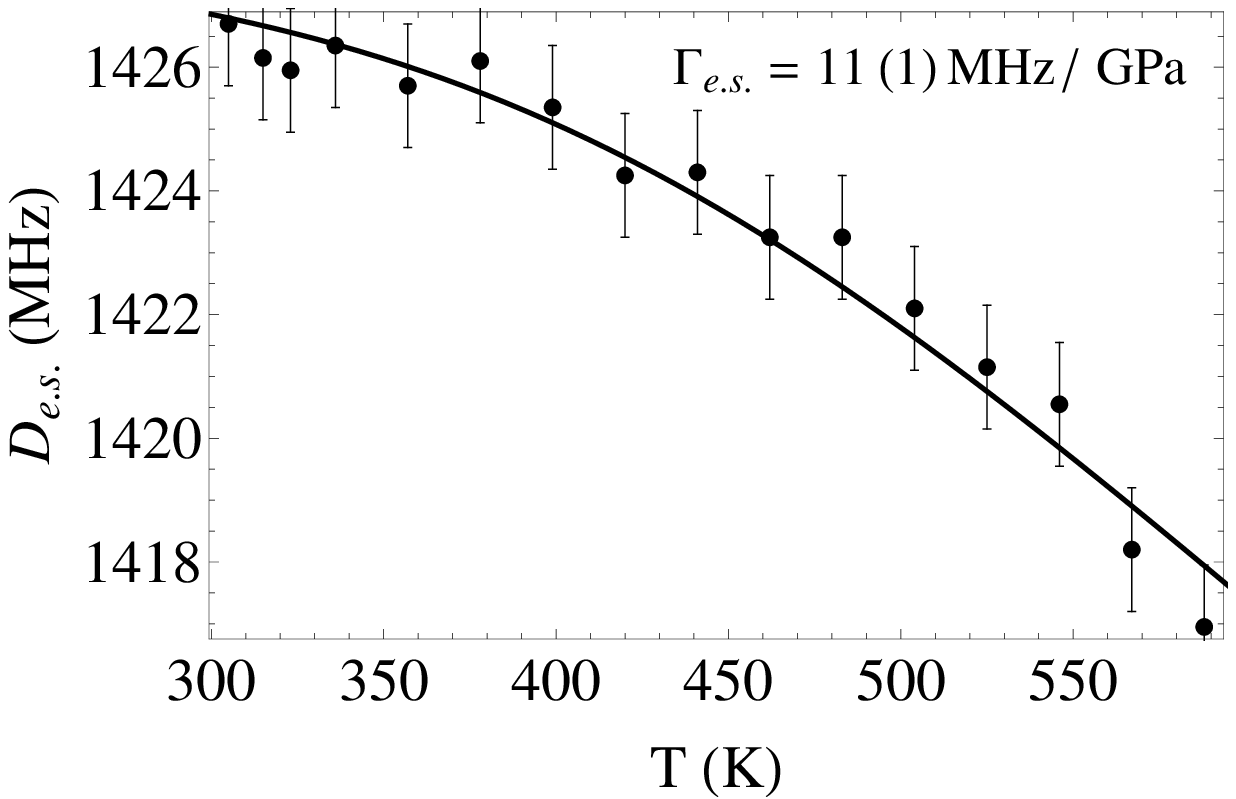}}}
\mbox{
\subfigure[]{\includegraphics[width=0.32\columnwidth] {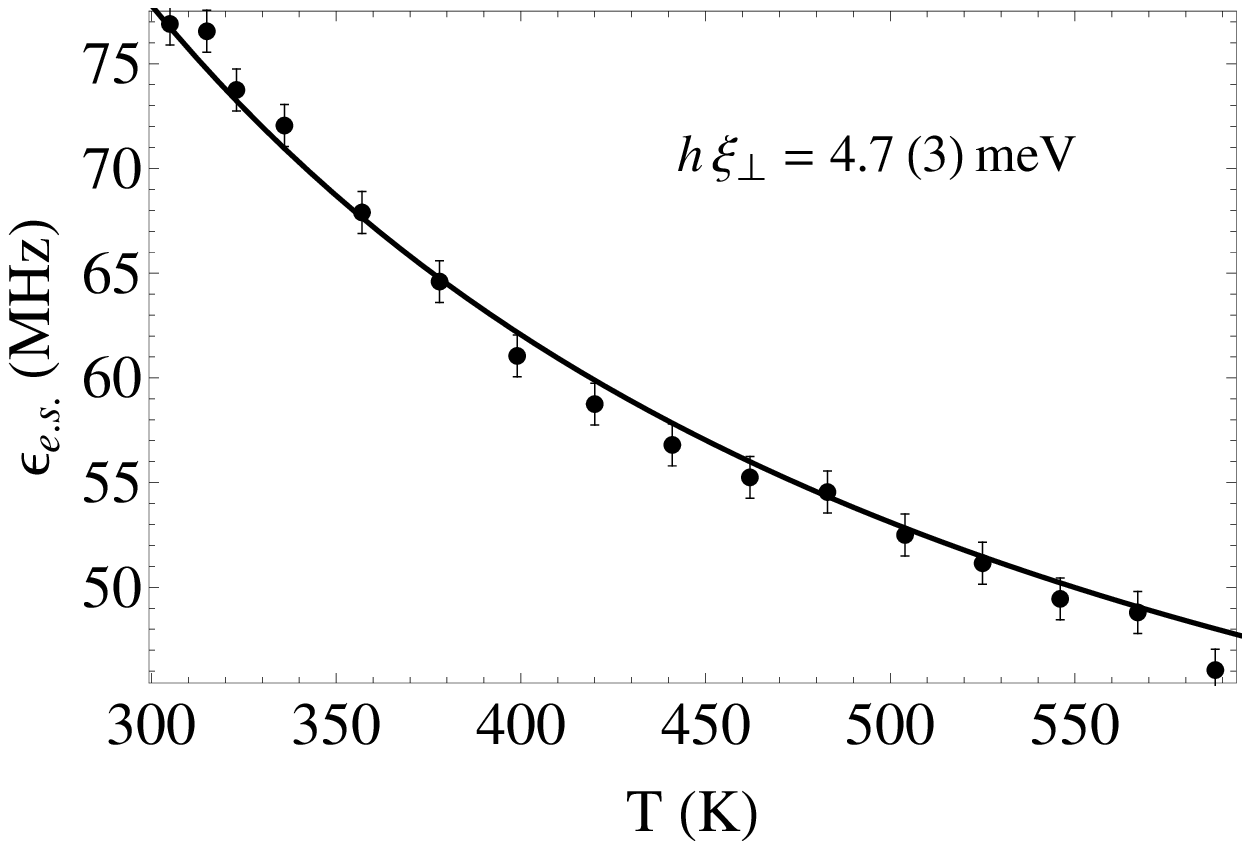}}
\subfigure[]{\includegraphics[width=0.33\columnwidth] {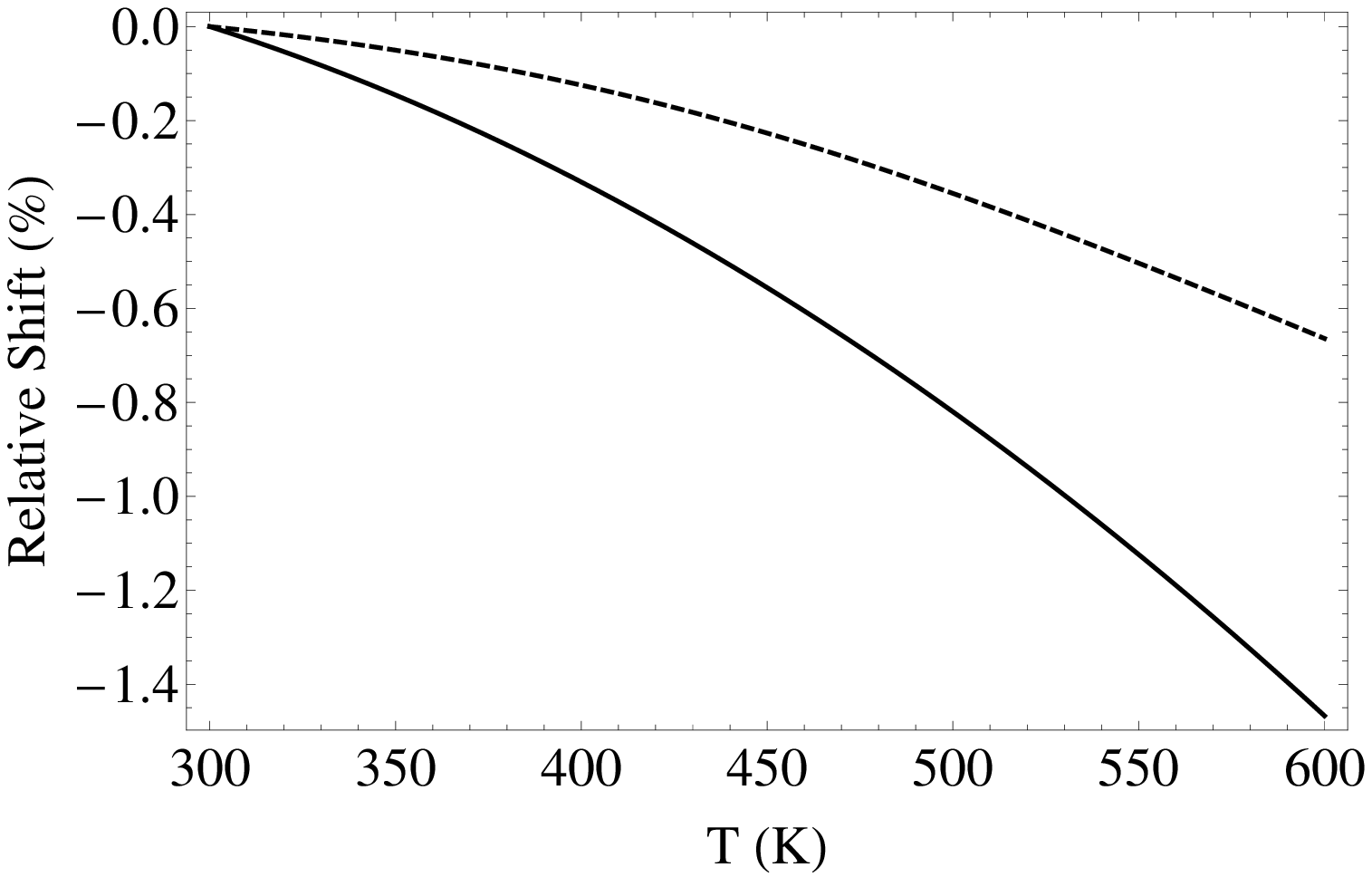}}
}
\caption{ (a) Example $^3E$ ODMR spectra at different temperatures. (b) The observed temperature variation of ${\cal D}_\mathrm{e.s.}$ and the fit of the thermal expansion contribution $\Delta {\cal D}_\mathrm{e.s.}^\mathrm{ex}(T)=\Gamma_\mathrm{e.s.}P(T)$. (c) The observed temperature dependence of ${\cal E}_\mathrm{e.s.}$ and the fit given by the expression (\ref{eq:splittingfitfunction}) with the single fit parameter $h\xi_\perp$. (d) Contrast of the relative temperature shifts of ${\cal D}_\mathrm{g.s.}$ (solid) and ${\cal D}_\mathrm{e.s.}$ (dashed), where relative shift is defined as $\left[{\cal D}_\alpha(T)-{\cal D}_\alpha(294 \mathrm{K})\right]/{\cal D}_\alpha(294 \mathrm{K})$ for $\alpha=\mathrm{g.s.}$ or $\mathrm{e.s.}$. In (b) and (c), the temperature error bars are smaller than the point size and the frequency errors bars are derived from the uncertainties of lorentzian lineshape fits of the ODMRs (see supplementary information).}
\label{fig:excitedstateshift}
\end{center}
\end{figure}

At each temperature, the observed $^3E$ ODMR spectra exhibited just two lines with no hyperfine structure (see Figure \ref{fig:excitedstateshift}a). This is due to the strain at this particular center being sufficiently large that the hyperfine resonances overlap (as depicted in Figure \ref{fig:electronicstructure}). In this regime, the minor splitting of the $m_s=\pm1$ sub-levels is the weighted average of the hyperfine resonances $2\epsilon_\mathrm{e.s.}(T)$, where
\begin{equation}
\epsilon_\mathrm{e.s.}(T) = \frac{1}{3}{\cal E}_\mathrm{e.s.}(T)+\frac{2}{3}\left[{\cal A}_\mathrm{e.s.}^{\parallel 2}+{\cal E}_\mathrm{e.s.}^2(T)\right]^{\frac{1}{2}}
\label{eq:splittingfitfunction}
\end{equation}
The observed temperature variation of ${\cal D}_\mathrm{e.s.}$ and $\epsilon_\mathrm{e.s.}$ are depicted in Figures \ref{fig:excitedstateshift}b and \ref{fig:excitedstateshift}c. As demonstrated in Figure \ref{fig:excitedstateshift}d, the relative shift of ${\cal D}_\mathrm{e.s}$ is much less than ${\cal D}_\mathrm{g.s.}$. Furthermore, as seen in Figure \ref{fig:excitedstateshift}b, the temperature shift of ${\cal D}_\mathrm{e.s.}$ is well described by just the contribution $\Delta {\cal D}_\mathrm{e.s.}^\mathrm{ex}(T)$ of thermal expansion with a hydrostatic pressure shift of $\Gamma_\mathrm{e.s.}=11(1)$ MHz/GPa. Given the observed $^3E$ ODMR linewidths, this shift indicates a projected pressure sensitivity of $\sim8(1)$ MPa $\mathrm{Hz}^{-1/2}$, which is comparable to the $\sim0.6$ MPa $\mathrm{Hz}^{-1/2}$ sensitivity of the ground state ODMR \cite{doherty14}. The absence of contributions of electron-phonon interactions is theoretically interesting and also explains why the relative shift of ${\cal D}_\mathrm{e.s.}$ is much smaller than ${\cal D}_\mathrm{g.s.}$. The implications of these conclusions for the first-principles theory of the temperature shifts is explored in the supplementary information.

As seen in Figure \ref{fig:excitedstateshift}c, a good fit of the observed temperature variation of $\epsilon_\mathrm{e.s.}(T)$ is obtained using the expression (\ref{eq:splittingfitfunction}). The only free parameter in the fit is the strain energy $h\xi_\perp=4.7(3)$ meV. This parameter is also the splitting of the visible ZPL, which was measured independently via room-temperature optical spectroscopy to be $\sim 4.3(2)$ meV. This small, one standard of deviation discrepancy between the independent measures, supports our model of the temperature dependence of the $^3E$ fine structure. The implication of the fit is that the temperature variation of ${\cal E}_\mathrm{e.s.}$ is primarily described by the strain-temperature reduction factor ${\cal R}(T)$, rather than the intrinsic temperature variation of the transverse spin-spin interaction $D_\mathrm{e.s.}^\perp$. Application of first-principles theory derives many similarities in the expected temperature variations of $D_\mathrm{e.s.}^\parallel$ and $D_\mathrm{e.s.}^\perp$ (see supplementary information). Consequently, the insignificant variation of $D_\mathrm{e.s.}^\perp$ is consistent with the observation that the variation of $D_\mathrm{e.s.}^\parallel$ is much smaller than $D_\mathrm{g.s.}^\parallel$ and is predominately due to thermal expansion.

This new knowledge of the fine, hyperfine and thermal properties of the excited state spin resonances provides the basis for a more comprehensive assessment of the resonances as additional quantum resources for the center's applications. For example, the excited state spin may be employed to perform fast, optically gated swap operations on a nuclear spin memory in ambient conditions \cite{fuchs10}. Alternatively, the differences in the properties of the ground and excited state spin resonances may be exploited to achieve bi-modal sensing, where the spins are conditioned to sense different effects concurrently. For example, the pressure sensitivities of the spin resonances of both states are comparable, but the  excited state spin resonances in the absence of stress are insensitive to temperature compared to the ground state spin resonances. Consequently, the superior temperature sensitivity of the ground state spin could be combined with the comparable pressure sensitivity of the excited state spin to sense pressure and temperature by iterating pulsed-ODMR sequences on the spins in rapid succession.

\begin{acknowledgement}
This work was supported by the Australian Research Council under the Discovery Project scheme DP0771676 and DP120102232.
\end{acknowledgement}

\begin{suppinfo}
Further theoretical and experimental details.
\end{suppinfo}

\end{document}